# Can Renewable Energy Mitigate Inflationary Pressures from Energy Imports? Evidence from Türkiye

Emre AKUSTA[1]

Yenilenebilir Enerji, Enerji İthalatından Kaynaklanan Enflasyonist Baskıları Azaltabilir mi? Türkiye'den Kanıtlar

*Öz*

Bu çalışma, yenilenebilir enerjinin Türkiye'de enerji ithalatından kaynaklanan enflasyonist baskıları azaltma potansiyelini analiz etmektedir. Çalışmada, birim kök analizi kapsamında yapısal kırılmaları dikkate alan Zivot & Andrews ile Lee & Strazicich testleri kullanılmıştır. Eşbütünleşme ilişkisi Johansen ve Hatemi-J eşbütünleşme testleri aracılığıyla incelenmiştir. Uzun dönem katsayıları DOLS ve FMOLS yöntemleri kullanılarak tahmin edilmiştir. Elde edilen bulguların sağlamlığı ise ayrıca ARDL yaklaşımı ile test edilmiştir. Ayrıca, enerji ithalatından kaynaklanan enflasyonist baskıları hafifletmede yenilenebilir enerjinin etkisini ölçmek için bir etkileşim terimi oluşturulmuştur. Sonuçlar, enerji ithalatı ve döviz kurunun enflasyon üzerinde artırıcı bir etkiye sahipken, yenilenebilir enerji ve etkileşim teriminin azaltıcı bir etkiye sahip olduğunu göstermektedir. DOLS, FMOLS ve ARDL sonuçları birbirini desteklemektedir. Ayrıca, her iki modelde de yenilenebilir enerjinin enerji ithalatından kaynaklanan enflasyonist baskıları hafifletme etkisi, yenilenebilir enerjinin doğrudan enflasyonu düşürücü etkisinden daha güçlüdür.

Can Renewable Energy Mitigate Inflationary Pressures from Energy Imports? Evidence from Türkiye

*Abstract*

This study analyses the potential of renewable energy to reduce inflationary pressures arising from energy imports in Türkiye. Annual data for the period 1980-2022 are used in the analysis. In this study, unit root properties are examined using the Zivot-Andrews and Lee-Strazicich tests, both of which explicitly account for structural breaks. Cointegration is investigated via the Johansen and Hatemi-J cointegration tests. Long-run coefficients are subsequently estimated using the DOLS and FMOLS estimators. The robustness of the empirical findings is further assessed using the ARDL approach. In addition, an interaction term is constructed to measure the impact of renewable energy in alleviating inflationary pressures arising from energy imports. The results show that energy imports and exchange rate have an increasing impact on inflation, while renewable energy and the interaction term have a decreasing impact. DOLS, FMOLS, and ARDL results support each other. Moreover, in both models, the impact of renewable energy in mitigating inflationary pressures stemming from energy imports is stronger than the direct disinflationary impact of renewable energy.

[1] Assoc. Prof. Dr., Kırklareli University, Faculty of Economics and Administrative Sciences, Department of Economics. E-mail: emre.akusta@klu.edu.tr. ORCID ID: https://orcid.org/0000-0002-6147-5443

## 1. Introduction

Energy is essential for modern economies. Energy plays a vital role in both industrial production and the sustainability of daily life. However, the unequal distribution of energy resources around the world leads to serious differences between countries. In particular, countries with insufficient energy resources are becoming dependent on foreign sources to meet their energy needs (Herbstreuth, 2014). Türkiye, as a country with limited resources in terms of energy production, meets most of its energy needs through imports. This situation causes Türkiye to be more affected by fluctuations in international energy markets and sudden changes in global energy prices. These changes in energy prices directly impact the prices in Türkiye's domestic market and create a serious pressure on inflation (Demirbas & Bakis, 2004; Uslu, 2008).

Energy imports have a significant impact on Türkiye's macroeconomic indicators. Especially the increase in energy prices is one of the most important factors triggering cost inflation. Energy is a basic input in many areas from industrial production to the service sector. Therefore, increases in energy prices are directly reflected in production costs. This leads to an increase in final product prices (Batten, Mo, & Pourkhanali, 2024). In countries with limited energy resources, such as Türkiye, the imbalance between energy demand and supply increases demand inflation. Dependence on energy imports, combined with exchange rate fluctuations, increases inflationary pressures.

The relationship between energy imports and exchange rates is particularly evident in developing countries. Türkiye is rapidly affected by exchange rate fluctuations due to its dependence on energy imports. This puts direct pressure on prices in the domestic market. In recent years, uncertainties and fluctuations in global markets have caused energy prices and exchange rates to fluctuate rapidly. These fluctuations have further triggered inflation in Türkiye's economic structure (Yorucu, 2016). Türkiye is one of the countries most affected by fluctuations in global energy prices due to its external energy dependence. Increases in energy prices directly increase production and transportation costs, which in turn are reflected in final product prices. Especially high oil prices increase transportation costs. As a result, prices of basic consumption goods, particularly food prices, have skyrocketed. This situation has severely reduced the purchasing power of households. As a result, additional measures had to be taken to fight inflation (Uslu, 2008). However, while such measures offer short-term solutions, more comprehensive structural steps need to be taken to reduce inflation permanently.

Renewable energy could be an important opportunity to reduce Türkiye's dependence on energy imports and alleviate inflationary pressures. The use of renewable energy sources can reduce external dependence by reducing energy imports. This can, in turn, alleviate pressure on the exchange rate and reduce inflationary impacts. Renewable energy can also contribute to Türkiye's long-term energy strategies by increasing environmental sustainability (Dogan & Muhammad, 2019; Katircioglu, Katircioglu, & Altinay, 2017). However, more research is needed on the economic consequences of this transition and its impacts on inflation. The costs of the transition to renewable energy, infrastructure investments and policy changes are factors that need to be considered for this process to be successful.

The potential of renewable energy to alleviate inflationary pressures from energy imports is an important research subject. Therefore, this study analyzes this potential for Türkiye using annual data for the period 1980-2022. Moreover, an interaction term is constructed to measure the impact of renewable energy on mitigating inflationary pressures from energy imports. The interaction term is used in statistical modeling to examine the joint impact of two or more independent variables on the dependent variable. With this method, in addition to the independent impacts of energy imports and renewable energy use on inflation, the combined impact of these two factors is also measured.

This study can contribute to the literature in at least five ways: (1) This study analyzes the potential of renewable energy to reduce inflationary pressures arising from energy imports. To the best of our knowledge, no empirical study has investigated this potential for Türkiye. This study aims to fill this gap in literature. (2) Unlike other studies, this study uses the interaction term. The interaction term aims to measure the independent impacts of energy imports and renewable energy use on inflation as well as the combined impact of these two factors. This allows for a more in-depth analysis. (3) Instead of traditional unit root tests, unit root tests that take structural breaks into account are used. This method allows for a better analysis of the dynamic structure of time series. It also increases the robustness of the results. (4) In addition to cointegration relationships, the study also estimates long-run coefficients. These estimates contribute to an in-depth analysis of the long-term impacts of energy imports and renewable energy use. (5) DOLS and FMOLS methods are used in the analysis and the results are compared. The use of these two methods improved the accuracy of the estimation results and strengthened the reliability of the analysis.

The rest of the paper is organized as follows: Section 2 reviews the literature, Section 3 describes the data and methodology, Section 4 presents the empirical findings, Section 5 reports the robustness checks, Section 6 contains policy implications, and finally Section 7 presents the conclusions.

## 2. Literature Review

This study examines the relationship between energy imports and inflation. It also analyzes the potential of renewable energy in Türkiye to reduce inflationary pressures from energy imports. Energy plays a central role for economies. This has continuously increased the number and scope of studies on energy. Therefore, the literature review was conducted in three stages. In the first stage, studies analyzing the impact of energy imports on macroeconomic indicators are discussed. In the second stage, studies analyzing the relationship between energy imports and inflation were evaluated. In the third stage, the literature analyzing the relationship between inflation and exchange rates is reviewed.

### 2.1. Research on Energy Imports and Macroeconomic Indicators

There are many studies in literature analysing the relationship between energy consumption, economic growth and other macroeconomic indicators. Among these studies, Asafu-Adjaye (2000) analysed the relationship between energy consumption, real income and consumer price index in Indonesia, India, Philippines and Thailand. In the analyses he conducted, he found that there is a reciprocal causality relationship between energy, income and consumer price index. Paul and Bhattacharya (2004) analysed the relationship between energy consumption and economic growth in India. Based on the data for the period 1950-1996, they found that there is a bidirectional causality relationship. Guttormsen (2007) analysed the relationship between energy consumption and economic growth in various countries such as France, Germany and Greece for the period 1960-2002. The results show that energy consumption and economic growth affect each other. Tsani (2010) analysed the relationship between energy consumption and economic growth in Greece on a sectoral basis. The results show that there is a unidirectional causality from total energy consumption to economic growth. However, there is no relationship between energy consumption in transport and economic growth.

Similar findings are also found in studies conducted in the Middle East and South American countries. Sadorsky (2011) analysed the relationship between energy consumption, economic growth and foreign trade in 8 Middle Eastern countries in the period 1980-2007. The results of the study show that there is a bidirectional causality between energy consumption and imports. Sadorsky (2012), meanwhile, found that there are both short-run and long-run relationships between energy consumption and foreign trade in South American countries. Aimer and Dilek (2021) analysed the relationship between energy consumption and economic growth in 16 Middle Eastern and North

African countries. The results revealed that there is a long-run relationship between economic growth and energy consumption.

In Türkiye-specific studies, Yanar and Kerimoğlu (2011) analysed the relationship between energy consumption, current account deficit and economic growth in Türkiye for the period 1975-2009. The results show that there is a long-run relationship between energy consumption and economic growth. Sancar and Polat (2015) also conducted a similar study for Türkiye. They found that there is unidirectional causality between energy consumption and economic growth and bidirectional causality between imports and energy consumption. Yılmaz and Altay (2016) analysed the relationship between energy consumption and inflation in Türkiye for the period 1970-2014. They found that the impact of energy consumption on inflation is negative in the long run. In addition, causality analyses revealed a causality relationship from inflation to energy consumption. Finally, Kızıldere (2020) analysed the impact of energy consumption and economic growth on the current account deficit in Türkiye for the period 1974-2015. The study found that these two variables create causality towards the current account deficit.

### 2.2. Research on Energy Imports and Inflation

Studies analyzing the relationship between energy imports and inflation evaluate this relationship in different countries in different periods. Among these studies, Ziaei (2012) analyzed the relationship between energy consumption, inflation, investment and public debt in 15 European countries. The study found a positive correlation between energy consumption and these variables. In addition, the study indicated that the major impact on energy consumption is due to investments. Iyke (2024) analyzed the relationship between energy consumption, economic growth and inflation in the case of Nigeria. Inflation was found to be the cause of energy consumption in the short run.

In studies conducted from a broader geographical perspective, the impacts of oil prices and energy consumption on inflation differ across countries. Abbas, Saeed, Manzoor, Arshad, and Bilal (2015) analysed the relationship between energy consumption, inflation and economic growth in Pakistan, China, India, Malaysia and South Africa. They concluded that there is no direct relationship between energy consumption and inflation in these countries. However, Yurtkur and Bahtiyar (2017) analysed Türkiye, India, Brazil, South Africa and Indonesia and found that shocks to economic growth positively impact energy consumption. However, they find that changes in energy consumption in Brazil, Türkiye and India are largely associated with fluctuations in economic growth. Alagöz, Alacahan, and Akarsu (2017) conducted a panel data analysis covering various countries such as Türkiye and China. They find that a one-unit increase in oil prices increases inflation by 0.04 units. Hammoudeh and Reboredo (2018) analysed the relationship between oil prices and inflation in the US. Since the US is an oil-exporting country, they concluded that the increase in oil prices has a different inflationary impact in the US compared to developing countries. In addition to oil prices, the relationship between energy consumption and inflation has also produced different results in various countries. Yakut, Yazgan, Bacaksız and Fikir (2021) find that there is no significant relationship between energy consumption and inflation in Türkiye and similar countries. This result indicates that the relationship between energy consumption and inflation may not be in the same direction for all countries. The impact of energy imports on inflation has also been extensively analyzed in the Türkiye perspective. Using a VAR model, Öksüzler and İpek (2011) finds that an increase in oil prices increases inflation in Türkiye. In a similar study, Yaylalı and Lebe (2012) analyzed the impact of oil prices on the Turkish economy. They find that imported oil prices have a causal impact on inflation. These two studies emphasize the role of oil price fluctuations on inflationary pressures in Türkiye. Kargi (2014) argues that oil prices are a cause of inflation in Türkiye. Moreover, it is supported by an empirical analysis that increases in oil prices significantly increase inflation in Türkiye.

### 2.3. Research on Inflation and Exchange Rate

There are many studies analyzing the relationship between exchange rates, inflation and interest rates. Ca' Zorzi, Hahn, and Sanchez (2007) analyzed 12 emerging economies and found that exchange rate pass-through to prices is significant. In the same period, Saha and Zhang (2013) analyzed the pass-through of the exchange rate to import, producer and consumer prices in Asia Pacific economies. The results show that the pass-through of the exchange rate to prices is higher in Australia than in China and India. Studies conducted in Africa and the Middle East show that the impact of exchange rates on inflation yields similar results in different geographies. Sheefeni and Ocran (2014) found that the exchange rate has a long-run causal impact on inflation in the Namibian economy. Helali and Rekik (2014) analyzed the direct impact of exchange rate on prices in Tunisian economy using SVAR and VECM methods. The results show that the exchange rate has a strong impact on prices. Syzdykova (2016) analyzed the relationship between nominal exchange rate and inflation in BRIC countries. A long-run relationship was found in all countries except China. Another study on Iran was conducted by Monfared and Akin (2017). The results of the study revealed that exchange rate and inflation have a linear relationship with each other. Moreover, the increase in money supply impacts inflation more than the exchange rate.

Studies analyzing the relationship between exchange rates, inflation and interest rates in Türkiye are also noteworthy. Sever and Mızrak (2007) analyzed the causal relationship between exchange rates, inflation and interest rates using VAR method. Data for the period 1987-2006 were used in the study. It is concluded that the increase in the exchange rate has a strong causal impact on inflation and interest rates. Similarly, Gül and Ekinci (2006) evaluated the relationship between exchange rate and inflation using Granger causality analysis. They find a unidirectional causality relationship from exchange rate to inflation. In Türkiye, Güneş (2013) analyzed the long-run interaction between exchange rate and inflation using cointegration and VECM methods. The study revealed that there is a causality relationship between exchange rate to inflation. Türk (2016) analyzed the relationship between exchange rate and inflation in Türkiye. In the study, it is stated that the exchange rate has a significant impact on inflation, but the impact of inflation on the exchange rate is not significant.

Yenice and Yenisu (2019) find a unidirectional causality relationship from exchange rates to inflation and interest rates using data for the period 2003-2018. Similarly, Duman and Sağdıç (2019) find that the increase in inflation is associated with changes in the real effective exchange rate. They argue that the Turkish economy should develop short and long-term policies to solve these problems. Among the recent studies, Demez (2021) stated that there is a long-run relationship between the exchange rate and inflation and that interest rates are effective on the exchange rate. Şeker (2022) found a causality relationship from exchange rate to inflation. Moreover, changes in exchange rates have a significant impact on inflation. Şanlı (2022) analyzed the impact of exchange rate and economic growth on inflation. The results show that a 1% increase in the exchange rate leads to a 0.5% increase in inflation. Konak and Peçe (2023) analyzed the relationship between exchange rate, interest rate and inflation in Türkiye for the period 2011-2021. As a result of the study, it was found that there is a cointegrated relationship between these variables in the long run. Moreover, there is a bidirectional causality between the exchange rate and inflation.

### 2.4. General Review on the Literature

There are various studies in the literature on the relationship between energy prices, exchange rates, inflation and energy consumption. Especially the impact of energy prices on inflation has been frequently analyzed by researchers. Studies such as İpek (2011), Yaylalı and Lebe (2012), Kargi (2014), Alagöz et al. (2017), and Hammoudeh and Reboredo (2018) show that increases in oil prices have a significant impact on inflation. These studies generally conclude that increases in energy prices increase inflationary pressures. Studies focusing on the relationship between exchange rates and inflation have also provided similar results. Studies such as Sever and Mızrak (2006), Gül and Ekinci (2006), Güneş (2013), Türk (2016), Monfared and Akın (2017), and Yenice and Yenisu (2019) have shown that changes in exchange rates have a significant impact on inflation. In these studies, a causality relationship from the exchange rate to inflation is generally found.

The relationship between energy consumption and economic growth is another area of intensive research. Studies such as Paul and Bhattacharya (2004), Guttormsen (2007), and Yanar and Kerimoğlu (2011) argue that there is an interaction between energy consumption and economic growth. These studies show that energy consumption moves together with economic growth. These two variables affect each other. Moreover, the relationship between energy consumption and macroeconomic variables such as imports and current account deficit has also been analyzed by many studies such as Sadorsky (2011, 2012), Sancar and Polat (2015). Studies examining the relationship between inflation and energy consumption have analyzed the impacts of energy consumption on inflation. Researchers such as Asafu-Adjaye (2000), Yılmaz and Altay (2016), and Hadi and Campbell (2020) analyzed the impact of energy consumption on inflation and concluded that energy prices generally increase inflationary pressures. These findings suggest that the relationship between energy consumption and inflation is an important issue that needs to be carefully analyzed.

In conclusion, there are many studies in literature analyzing the relationship between energy prices, exchange rates, energy consumption and inflation. However, this study makes a different contribution to literature. This study also analyzes the potential of renewable energy in Türkiye to reduce inflationary pressures arising from energy imports. Most of the studies in the literature analyzing the relationship between energy prices, consumption and inflation do not focus on renewable energy. The impacts of renewable energy on inflation in an energy import-dependent economy like Türkiye are evaluated. In this regard, this study is expected to make an important contribution to literature.

## 3. Data and Methodology

### 3.1. Model Specification and Data

The empirical analysis of this study examines whether renewable energy can alleviate inflationary pressures stemming from energy imports in the case of Türkiye. The study uses annual data for the period 1980-2022. This period was chosen based on the availability and suitability of the dataset for the analysis. Inflation (INF) is used as the dependent variable. Energy imports (IMP), renewable energy use (REN) and exchange rate (EXC) are included in the model as independent variables. In addition, an interaction term[2] is constructed to measure the impact of renewable energy in mitigating inflationary pressures from energy imports. The interaction term is a method used in statistical modeling to examine the combined impact of two or more independent variables on the dependent variable (Adedoyin, Ozturk, Bekun, Agboola, & Agboola, 2021; Hossain et al., 2022). This term, especially in multivariate regression analyses, is used to show how independent variables interact with each other to create an interaction beyond their direct impact on the dependent variable. In other words, it measures the independent impacts of energy imports and renewable energy use on inflation as well as the combined impact of these two factors.

[2] The interaction term has been constructed as $REN \times IMP$. A statistically significant negative coefficient on the interaction term indicates that higher renewable energy use mitigates the inflationary effect of energy imports, and the opposite case is also valid.

The model of the research is given in the functional form in Equation 1:

$$INF_t = \beta_0 + \beta_1 IMP_t + \beta_2 REN_t + \beta_3 EXC_t + \beta_4 TERM_t + \epsilon_t \quad (1)$$

In this model, $\beta_0$ represents the constant term, while the coefficients $\beta_1$ to $\beta_4$ measure the impact of each independent variable on exports. $\epsilon_t$ is the error term with zero mean and constant variance and $t$ is the time interval.

The indicators used in our study and their descriptive statistics are shown in Table 1.

**Table 1.** Descriptive statistics

| Variables | Symbol | Description | Obs. | Mean | SD | Min. | Max. | Source |
|---|---|---|---|---|---|---|---|---|
| Inflation | INF | Consumer prices (%) | 39.39 | 29.89 | 6.251 | 105.22 | 39.39 | WB |
| Energy imports | IMP | Net (% of energy use) | 62.1 | 11.71 | 43.04 | 76.285 | 62.1 | EIA |
| Renewable energy | REN | Net (% of energy use) | 19.88 | 6.491 | 8.475 | 35.734 | 19.88 | EIA |
| Exchange rate | EXC | Dollar exchange rate | 1.762 | 3.016 | 0.001 | 16.594 | 1.762 | EVDS |

Note: (1) WB, EIA, and EVDS indicate World Bank-World Development Indicators, U.S. Energy Information Administration-International Energy Statistics database, and Central Bank of the Republic of Türkiye-EVDS data central, respectively. (2) The abbreviations N, Obs, SE, Min and Max denote the number of observations, standard deviation, standard error, minimum and maximum values, respectively.

Table 1 shows the descriptive statistics of the variables used in the study. Since the variables used are proportional and percentage, their logarithms are not taken. INF denotes inflation. The average value of INF is calculated as 39.39%. With a standard deviation of 29.89%, inflation experienced significant fluctuations throughout the period. The lowest value was recorded as 6.25% and the highest value as 105.22%. This shows that inflation in Türkiye was quite high in some years. IMP shows the share of imported energy in energy use. The average IMP is 62.1%. This means that a large portion of Türkiye's energy needs are met by imports. With a standard deviation of 11.71%, energy imports range from a minimum of 43.04% to a maximum of 76.285%. REN refers to the portion of energy use derived from renewable resources. With an average of 19.88%, this variable shows that the contribution of renewable energy to Türkiye's energy production is limited. However, it varied between a minimum of 8.475% and a maximum of 35.734% over the period. EXC represents the value of the Turkish Lira against the dollar. With a mean value of 1.762, the exchange rate fluctuated widely with a standard deviation of 3.016. The lowest exchange rate value was recorded at 0.001 and the highest at 16.594. This reflects the massive exchange rate movements in Türkiye over the period.

### 3.2. Unit Root Analysis

Stationarity means that the mean, variance and autocorrelation of the series remain constant over time. Models may give biased results in non-stationary series (Granger & Newbold, 1974). Therefore, unit root tests are used to determine whether a time series is stationary. Series with unit roots are non-stationary and exhibit a random walk. Structural breaks or exogenous shocks are common in economic and financial series (Johansen & Juselius, 1990). In such cases, conventional unit root tests that assume a stable data-generating process may yield misleading inferences. The Zivot and Andrews (1992) test was developed to solve this problem. This test provides more accurate results by considering possible structural breaks in the series. Similarly, the Lee and Strazicich (2003) test extends the analysis by allowing for endogenous breaks within an LM testing framework. Breaks can be caused by events such as economic crises, policy changes or sudden shocks.

*The Zivot and Andrews (1992)* test was applied in this study to evaluate the stationarity properties of time series. Since traditional unit root tests assume fixed structures in data series, they may ignore possible structural breaks in the series and may produce misleading results. The Zivot & Andrews test, on the other hand, aims to improve the accuracy of the stationarity test by endogenously detecting a single structural break that may occur in the time series. Accordingly, the test is implemented under three specifications: Model A allows a break in the intercept (level), Model B allows a break in the trend, and Model C allows breaks in both the intercept and the trend (Zivot & Andrews, 1992).

*The Lee and Strazicich* (2003) LM unit root test is additionally employed as a complementary approach that allows for one or two endogenous structural breaks in the level and/or trend. A key advantage of this methodology is that breaks are permitted under both the null and the alternative hypotheses, which mitigates spurious rejections that may arise in break-augmented tests. Break dates are selected endogenously by searching over admissible partitions and choosing the specification that yields the most negative (minimum) LM test statistic, and inference is then conducted using the corresponding critical values.

### 3.3. Cointegration Test

In this study, Johansen (1988) and Hatemi-J (2008) cointegration test is employed to determine the long-run relationships among variables. Given that long-run relationships may be affected by regime shifts, policy changes, or crisis episodes, applying a break-augmented approach alongside a standard system-based test helps to strengthen the robustness of inference.

*The Johansen (1988) cointegration test* builds on the cointegration concept introduced by Engle and Granger (1987). It is designed for a multivariate setting. It tests whether multiple time series share a common long-run equilibrium relationship. The Johansen procedure is estimated via maximum likelihood and relies primarily on two statistics: the Trace test and the Maximum Eigenvalue test. The Trace test evaluates cointegration by starting from the null hypothesis of no cointegration and then proceeding sequentially to assess whether additional cointegrating relations exist in the system. The Maximum Eigenvalue test complements this by focusing on the incremental contribution of the next potential cointegrating relation (e.g., comparing the hypothesis of no cointegration against the hypothesis of at least one cointegrating relation), based on the largest eigenvalue (Johansen, 1988).

*The Hatemi-J (2008) cointegration test* is additionally applied to avoid overlooking potential structural breaks in the long-run relationship. Hatemi-J (2008) extends conventional cointegration testing by allowing for up to two endogenous structural breaks in the cointegration equation, which is particularly relevant in the presence of regime shifts, policy changes, or crisis periods. In this approach, break dates are not imposed a priori; instead, they are determined endogenously by searching over admissible break combinations and selecting the dates that provide the strongest evidence against the null of no cointegration. The resulting test statistic is then compared with the Hatemi-J (2008) critical values to evaluate whether cointegration holds once structural changes in the intercept and/or slope are considered.

### 3.4. Long-Run Coefficient Estimates

In this study, the Dynamic Ordinary Least Squares (DOLS) and Fully Modified Ordinary Least Squares (FMOLS) methods were used to estimate long-run coefficients. Additionally, the Autoregressive Distributed Lag (ARDL) method was used for robustness checks. These methods are powerful tools for estimating cointegration vectors. They are characterized by efficient estimation, especially for small sample sizes.

*The DOLS method* was developed by Saikkonen (1991) and Stock and Watson (1993). DOLS addresses the problems of autocorrelation and possible endogeneity by adding the differences and lagged terms of the series to the model when estimating the cointegration relationship. This method provides a robust estimation of the long-run coefficients with the added difference terms and future values.

*The FMOLS method* was developed to correct for possible autocorrelation and small sample bias when estimating cointegration vectors. This method was proposed by Phillips and Hansen (1990). It provides more accurate and consistent estimates by addressing endogeneity and autocorrelation in the cointegration relationship. The FMOLS model is more complex than the traditional ordinary least squares (OLS) method and corrects for bias in the estimates. FMOLS corrects for autocorrelation in the error term. It reveals the cointegration relationship in the series more reliably. Moreover, FMOLS improves coefficient consistency by eliminating potential endogeneity problems.

*The ARDL method* was developed by Pesaran and Shin (1999) and later refined through the bounds testing approach by Pesaran, Shin, and Smith (2001). It is a single-equation framework that captures both short-run dynamics and long-run relationships by including lags of the dependent and explanatory variables. A key advantage of ARDL is its applicability when regressors are a mix of I(0) and I(1), provided none are I(2). The existence of a long-run relationship is typically assessed using the bounds test based on the joint significance of lagged level terms. Once cointegration is supported, the model is re-parameterized into an error-correction form to estimate long-run coefficients and short-run adjustments. Lag orders are generally selected using standard information criteria.

In this study, long-run relationships have been estimated using multiple forecasting methods. These forecasters are widely used in cointegration settings to reduce endogeneity and serial correlation issues and improve small sample performance. The main discussion is based on the implied long-run coefficients. Moreover, since FMOLS may become problematic when the regressors are cointegrated among themselves, FMOLS results are interpreted with caution and greater weight is placed on DOLS/ARDL-based evidence, consistent with Yahyaoui and Bouchoucha (2021). The findings from both approaches are reported comparatively.

**4. Empirical Findings**

Prior to the cointegration analysis and the estimation of long-run coefficients, the stationarity properties of the series were examined. To account for potential structural changes, unit root testing was conducted using the Zivot & Andrews and Lee & Strazicich structural-break unit root tests. The unit root test results are reported in Table 2 and Table 3.

**Table 2.** Zivot & Andrews structural break unit root test results

| Variable | Level | | First difference | | Model |
|---|---|---|---|---|---|
| | Breakpoints | t-Statistic | Breakpoints | t-Statistic | |
| INF | 2003 | -3.818 | 1995 | -6.953*** | Model A |
| INF | 2002 | -3.471 | 2002 | -6.710*** | Model C |
| IMP | 1995 | -3.466 | 2006 | -7.862*** | Model A |
| IMP | 2000 | -3.332 | 2004 | -7.816*** | Model C |
| REN | 2015 | -4.482 | 2015 | -8.870*** | Model A |
| REN | 2007 | -4.802 | 2007 | -9.070*** | Model C |
| EXC | 2015 | -4.330 | 2002 | -6.120*** | Model A |
| EXC | 2016 | -2.799 | 2016 | -7.756*** | Model C |
| Critical value | Model A | → | %10: -4.58 | %5: -4.93 | %1: -5.34 |
| | Model C | → | %10: -4.82 | %5: -5.08 | %1: -5.57 |

Note: The superscripts ***, **, and * denote the significance at a 1%, 5%, and 10% level, respectively.

**Table 3.** Lee & Strazicich structural break unit root test results

| Variable | Level | | First difference | | Model |
|---|---|---|---|---|---|
| | Breakpoints | t-Statistic | Breakpoints | t-Statistic | |
| INF | (1988, 2003) | -2.902 | (1990, 1994) | -3.864** | Model A |
| INF | (1993, 2010) | -4.999* | (1985, 1994) | -6.539*** | Model C |
| IMP | (2008, 2012) | -3.294 | (2001, 2008) | -5.834*** | Model A |
| IMP | (1989, 2008) | -4.944 | (1999, 2009) | -7.632*** | Model C |
| REN | (2001, 2015) | -3.806* | (1988, 2010) | -9.495*** | Model A |
| REN | (1987, 2007) | -5.020* | (1987, 2010) | -9.643*** | Model C |
| EXC | (1985, 1990) | -0.324 | (2001, 2018) | -5.068*** | Model A |
| EXC | (1995, 2011) | -3.720 | (2007, 2018) | -7.316*** | Model C |
| Critical value | Model A | → | %10: -3.504 | %5: -3.842 | %1: -4.545 |
| | Model C | → | %10: -4.989 | %5: -5.286 | %1: -5.823 |

Note: The superscripts ***, **, and * denote the significance at a 1%, 5%, and 10% level, respectively.

Table 3 shows that INF is not stationary at level. However, it became stationary at first difference with the structural breaks in (1990, 1994) under Model A and (1985, 1994) under Model C. These breaks, particularly the 1994 break, coincide with a major crisis in Türkiye's economic history. The 1994 economic crisis led to a rapid depreciation of the Turkish Lira and a rapid increase in inflation. The additional break in the early-1990s window (captured as 1990/1985 depending on the specification) can be read as reflecting the build-up to, and immediate adjustment around, this episode. In addition, the level specifications identify breaks around 2003 (Model A) and 2010 (Model C), which are consistent with the post-2001 disinflation period and subsequent macro-financial regime shifts. The year 2003 coincides with the immediate aftermath of the 2001 economic crisis in Türkiye. The 2001 crisis was a serious crisis triggered by problems in the banking sector and fluctuations in exchange rates. During this period, inflation increased rapidly. However, with the economic reforms implemented after the crisis and the transition to the floating exchange rate regime, inflation started to be brought under control (Özlale & Yeldan, 2004; Atasoy & Saxena, 2006; Ardan, 2023).

Also, the Lee & Strazicich test indicates that IMP becomes stationary at first difference, implying that shocks to energy imports have persistent effects in levels. The break points for energy imports are determined as (2001, 2008) under Model A and (1999, 2009) under Model C. These periods are closely related to Türkiye's energy dependence and pronounced shifts in global energy markets. The early-2000s break (around 1999-2001) is consistent with the domestic crisis environment and subsequent policy reorientation, while the 2008-2009 breaks align with the global financial crisis period and the associated sharp swings in energy prices and external demand. These breaking points support the view that energy imports are sensitive to both global energy shocks and Türkiye's evolving energy demand (Celik, 2006; Hepbasli, 2005; Kucukali, 2010; Kucukali & Baris, 2009).

The results of the Lee & Strazicich test show that REN is weakly stationary at level once two breaks are allowed (with breaks in (2001, 2015) for Model A and (1987, 2007) for Model C), and it becomes clearly stationary at first difference with the breaks in (1988, 2010) for Model A and (1987, 2010) for Model C. The year 2007 was a time when the global interest in renewable energy started to increase. In parallel with this global trend, Türkiye also accelerated its renewable energy investments. 2015 was a period in which incentives and regulations for renewable energy intensified in Türkiye. In this period, investments in renewable resources such as solar and wind energy increased. Therefore, renewable energy has started to play an important role in Türkiye's energy policies. The additional breaks around 2010 and the earlier regime-shift dates (late-1980s/early-2000s, depending on the model) suggest that the renewables series reflects both the later acceleration phase and earlier structural changes that shaped the long-run trajectory of renewable energy in Türkiye. This result reveals that the shift towards renewable energy in Türkiye's energy policies has been evident since 2007 in line with global trends (Salvarli & Salvarli, 2017; Yüksel, 2008).

Moreover, the results show that EXC is not stationary at level, but becomes stationary at first difference with the breaks in (2001, 2018) under Model A and (2007, 2018) under Model C. The year 2001 is the period when Türkiye switched to a floating exchange rate regime after the 2001 crisis. This period represents significant changes in the foreign exchange market. This transition resulted in the exchange rate being determined under free market conditions. As a result, the value of the Turkish Lira fluctuated significantly. The 2018 break is consistent with the pronounced exchange-rate turbulence and sharp depreciation observed in that period, while the 2007 break can be interpreted as capturing pre-global-crisis macro-financial conditions and the ensuing shift in external financing dynamics. These findings suggest that structural changes in the exchange rate are closely related to Türkiye's economic reforms and global market conditions (Akyurek, 2006; Dağlaroğlu, Demirel, & Mahmud, 2018; Özlale & Yeldan, 2004).

Overall, the Lee & Strazicich results are broadly consistent with your earlier Zivot & Andrews interpretation, in that the series are predominantly non-stationary in levels but become stationary after first differencing once structural breaks are accounted for (with some variables showing only weak evidence of level stationarity under specific model choices).

**Table 4.** Johansen cointegration test results

| $H_0$ | Trace statistics | Critical values (5%) | Maximum eigenvalue statistics | Critical values (5%) |
|---|---|---|---|---|
| r = 0 | 53.586 | 47.856*** | 39.815 | 27.584*** |
| r ≤ 1 | 33.770 | 29.797*** | 44.508 | 21.131** |
| r ≤ 2 | 19.262 | 15.495*** | 18.455 | 14.265** |
| r ≤ 3 | 6.807 | 3.841*** | 6.807 | 3.841*** |

Note: The superscripts ***, **, and * denote the significance at a 1%, 5%, and 10% level, respectively.

Table 4 shows the results of the Johansen cointegration test. This test uses the Trace and Maximum Eigenvalue statistics to determine the existence of cointegration vectors between variables. First, with respect to the null hypothesis r = 0 (no cointegration), the Trace test statistic is 53.586 and the critical value is 47.856. The Maximum Eigenvalue statistic is 39.815, which is above the critical value (27.584). This result indicates that the null hypothesis should be rejected and there is at least one cointegration vector. Second, when the hypothesis r ≤ 1 is tested, the Trace test statistic is 33.770, which exceeds the critical value of 29.797. The Maximum Eigenvalue statistic is 44.508, which is higher than the critical value of 21.131. This indicates the existence of two cointegration vectors. Third, the Trace statistic for the hypothesis r ≤ 2 is 19.262, which is above the critical value (15.495). The Maximum Eigenvalue statistic is 18.455 and exceeds the critical value (14.265). This indicates that there are three cointegration vectors. Finally, according to the hypothesis r ≤ 3, both the Trace and Maximum Eigenvalue statistics are 6.807 and exceed the critical value (3.841). This indicates the existence of a fourth cointegration vector.

**Table 5.** Hatemi-J cointegration test results

| Test | t-Statistic | Breakpoints | Critical value | | |
|---|---|---|---|---|---|
| | | | 1% | 5% | 10% |
| ADF | -7.198*** | (1990, 2001) | -6.503 | -6.015 | -5.653 |
| Zt | -9.458*** | (2001, 2010) | -6.503 | -6.015 | -5.653 |
| Za | -89.178** | (1987, 2008) | -90.794 | -76.003 | -52.232 |

Note: The superscripts ***, **, and * denote the significance at a 1%, 5%, and 10% level, respectively.

Table 5 reports the results of the Hatemi-J cointegration test, which examines the existence of a long-run relationship among the variables while allowing for two endogenous structural breakpoints. For the ADF statistic, the test value is −7.198 with breakpoints at (1990, 2001). Since this statistic is lower than the 1% critical value (−6.503), the null hypothesis of no cointegration is rejected at the 1% significance level. Similarly, the Zt statistic is −9.458 with breakpoints at (2001, 2010), and it is also more negative than the 1% critical value (−6.503), providing further evidence against the null of no cointegration. Finally, the Za statistic is −89.178 with breakpoints at (1987, 2008). This value is more negative than the 5% critical value (−76.003), but not more negative than the 1% critical value (−90.794), indicating rejection of the null at the 5% level. Overall, the Hatemi-J test results strongly support the presence of cointegration, implying a robust long-run equilibrium relationship even after accounting for two structural breaks.

Overall, the empirical evidence is consistent across both cointegration approaches. Specifically, the Johansen test results indicate the presence of cointegration vectors among the variables, while the Hatemi-J test statistics (ADF, Zt, and Za) similarly reject the null hypothesis of no cointegration when structural breaks are endogenously taken into account. Hence, based on both the Johansen and Hatemi-J cointegration tests, it can be concluded that a stable long-run equilibrium relationship exists among the variables.

**Table 5.** DOLS and FMOLS long-run estimates

| Dependent variable: INF | Coefficient | |
|---|---|---|
| Variables | **DOLS** | **FMOLS** |
| IMP | 1.642*** | 1.188*** |
| REN | -1.145** | -1.983*** |
| EXC | 3.228*** | 3.702** |
| TERM | -1.346** | -2.193*** |
| C | -12.731** | -7.230*** |
| Diagnostic tests | P value | P value |
| $\chi^2$ (Serial correlation) | 0.314 | 0.237 |
| $\chi^2$ (Heteroskedasticity) | 0.672 | 0.589 |
| $\chi^2$ (Normality) | 0.934 | 0.764 |
| $\chi^2$ (Functional form) | 0.221 | 0.311 |
| CUSUM | Stable | Stable |
| CUSUMSQ | Stable | Stable |

Note: The superscripts ***, **, and * denote the significance at a 1%, 5%, and 10% level, respectively.

Table 5 presents both DOLS and FMOLS results. DOLS results show that a one-unit increase in the exchange rate (EXC) increases the inflation rate by 3.228 units. This result is quite significant for an import-dependent economy such as Türkiye's. An increase in the exchange rate increases import costs. Especially in energy import-dependent countries like Türkiye, increases in exchange rates cause inflationary pressures by increasing energy costs. Türkiye's energy and raw material imports directly affect the inflation. In this case, exchange rate increases lead to a significant pressure on inflation. Similarly, energy imports (IMP) also have an inflationary impact. The DOLS model reveals that a one-unit increase in energy imports increases inflation by 1,642 units. Türkiye's energy needs are largely met by imports. Increases in the prices of energy resources such as oil and natural gas trigger inflation by increasing the cost of energy imports. These factors increase Türkiye's sensitivity to fluctuations in global energy prices. It shows that Türkiye is vulnerable to exogenous shocks due to its energy dependence.

Renewable energy use (REN) has a mitigating impact on inflation. DOLS results show that a one-unit increase in REN reduces inflation by 1.145 units. This shows that reducing Türkiye's energy imports has a beneficial impact on inflation. Renewable energy sources play an important role in controlling inflation by reducing energy costs through the use of domestic resources. This result reveals the stabilizing effect of renewable energy investments on energy prices. Finally, the TERM variable refers to the interaction term. This term measures the impact of renewable energy on mitigating inflationary pressures from energy imports. The DOLS model shows that the interaction term has a negative effect. A one-unit increase in TERM reduces the inflation rate by 1.346 units. This result reveals that renewable energy mitigates inflationary pressures stemming from energy imports. Renewable energy reduces energy costs by reducing the use of imported fossil fuels. Thus, the pressure of energy imports on prices decreases. This shows how important Türkiye's investments in renewable energy are in controlling inflation. The findings of the study are consistent with studies that conclude that energy imports increase inflationary pressures (see, for example, Yılmaz & Altay, 2016; Hadi & Campbell, 2020). It is also consistent with studies that find that the relationship between exchange rate and inflation is significant (see, for example, Sever & Mızrak, 2006; Gül et al. 2006; Güneş, 2013; Türk, 2016; Monfared & Akın, 2017; Yenice & Yenisu, 2019).

FMOLS results also support the DOLS results. In both models, renewable energy sources have a disinflationary impact. Moreover, in both models, the impact of renewable energy on mitigating inflationary pressures stemming from energy imports is stronger than the direct disinflationary impact of renewable energy. This suggests that renewable energy investments will play a critical role in containing inflation in Türkiye in the long run. It is evident that renewable energy has the potential to reduce energy costs and stabilize inflation by reducing external dependence. These results indicate that shifting Türkiye's energy policy towards renewable energy is of strategic importance for both the economy and inflation management.

## 5. Robustness Check

To further assess robustness, additional checks have been conducted for both (i) cointegration inference and (ii) long-run coefficient estimates. Cointegration has been re-tested under endogenous structural breaks to avoid potentially biased conclusions in the presence of regime shifts. The Hatemi-J test results are reported in Table 6. Long-run and short-run dynamics have also been re-estimated within an ARDL error-correction framework. The ARDL short-run and long-run results are presented in Table 7.

**Table 6.** ARDL long-run estimates

| Dependent variable: INF | Coefficient | |
|---|---|---|
| Variables | **Short-run** | **Long-run** |
| IMP | 2.548*** | 2.429*** |
| REN | -2.209* | -2.504*** |
| EXC | 2.876*** | 3.237** |
| TERM | -2.203 | -2.999** |
| C | | -3.526** |
| Diagnostic tests | | P value |
| $\chi^2$ (Serial correlation) | | 0.247 |
| $\chi^2$ (Heteroskedasticity) | | 0.443 |
| $\chi^2$ (Normality) | | 0.349 |
| $\chi^2$ (Functional form) | | 0.121 |
| CUSUM | | Stable |
| CUSUMSQ | | Stable |

Note: The superscripts ***, **, and * denote the significance at a 1%, 5%, and 10% level, respectively.

Table 6 reports the ARDL estimates as a robustness check by decomposing the relationship into short-run dynamics and long-run equilibrium effects. The ARDL long-run coefficients are broadly consistent with the DOLS and FMOLS findings in Table 5. In particular, the exchange rate (EXC) remains inflationary and statistically significant. Energy imports (IMP) also continue to exert a positive and significant effect on inflation. These results confirm the relevance of external cost channels for an import-dependent economy such as Türkiye. Renewable energy (REN) retains a negative and statistically significant long-run coefficient. This implies that higher renewable energy utilization is associated with lower inflation by reducing reliance on imported energy inputs. With respect to the interaction term (TERM), it is constructed to capture the role of renewable energy in mitigating inflationary pressures arising from energy imports. The ARDL results indicate that TERM is statistically insignificant in the short run. However, it becomes negative and significant in the long run. This pattern suggests that the buffering role of renewable energy materializes mainly through longer-horizon adjustments, such as capacity expansion, substitution away from imported fossil fuels, and gradual cost pass-through. Finally, diagnostic statistics and stability tests indicate no major specification concerns. They also support parameter stability. Overall, the evidence reinforces the conclusion that the core results are robust across alternative long-run estimators and within an ARDL framework.

## 6. Policy Implications

The analysis results demonstrate that renewable energy resources can mitigate the inflationary pressures stemming from energy imports in Türkiye. In line with these findings, shifting Türkiye's energy policy towards renewable energy is of strategic importance for both economic and inflation management. Renewable energy can alleviate the pressure on energy costs by reducing dependence on energy imports. This would provide Türkiye with a more sustainable solution in the fight against inflation. Moreover, renewable energy sources not only ensure economic stability but also support environmental sustainability. Renewable energy plays an important role in stabilizing energy prices by reducing external dependence. In this respect, a number of policy recommendations have been developed to reduce dependence on energy imports and combat inflation:

[1] Promote renewable energy use and investments: Türkiye should develop a comprehensive policy package for this purpose. This package should include incentives such as tax breaks, subsidies, low-interest loans and direct investment support. Tax exemptions and customs duty reductions could be applied to companies investing in renewable energy projects. In addition, domestic production of technological equipment used in power generation plants should be encouraged. The government should offer grants to support R&D efforts. It should reduce capital costs by providing low-interest loans for renewable energy investments. Establish a long-term and stable regulatory framework to boost investor confidence. Price guarantees to energy producers should be stabilized and secured.

[2] Long-term strategies to reduce energy dependence: Türkiye should adopt long-term energy policies to reduce dependence on energy imports. Projects to increase energy efficiency should be prioritized. Policies that encourage energy savings in the public and private sectors should be implemented. In addition, investments in energy storage systems should be increased. Given the intermittent generation structure of renewable energy sources, these systems are critical for ensuring energy supply security.

[3] Transformation and incentive programs in energy-intensive sectors: Türkiye should implement programs to increase the use of renewable energy in energy-intensive sectors such as industry, agriculture and transportation. Policies should be developed to encourage the transition to renewable energy to reduce fossil fuel use. Tax reductions and incentives should be provided to support the use of energy efficient technologies. Reducing energy costs, especially in the industrial sector, will also reduce production costs and alleviate inflationary pressures.

[4] Establish a mechanism to monitor the relationship between energy and inflation: Türkiye should establish a mechanism that regularly monitors and evaluates the impacts of energy use on inflation. This mechanism would measure the short and long-term impacts of energy policies and allow for policy adjustments when necessary. Thus, flexible and sustainable solutions can be developed to combat energy market volatility. Such monitoring and evaluation systems are important to minimize the impact of energy prices on inflation. This will help policymakers make more informed decisions.

[5] Supporting domestic production in renewable energy technology: Importing technologies to be used in renewable energy power plants is a problem that increases Türkiye's external dependency. Supporting domestic production in renewable energy investments is critical for reducing Türkiye's external dependence in the energy sector. Domestic production of these technologies should be encouraged. The state should offer tax advantages, grant programs and low-interest loans to the private sector for domestic technology development and production. In addition, R&D efforts should be supported. Domestic producers' access to renewable energy technologies should be increased. These steps would lower energy costs, reduce Türkiye's dependence on energy imports and ease inflationary pressures. This would reduce import dependency in both raw materials and technology. As a result, a stronger domestic infrastructure will be built in the energy sector.

## 7. Conclusion

This study analyzes the potential of renewable energy to mitigate inflationary pressures stemming from energy imports in Türkiye. Annual data for the period 1980-2022 are used in the study. Unit root properties are examined using the Zivot-Andrews and Lee-Strazicich tests, both of which explicitly account for structural breaks. Cointegration is investigated via the Johansen and Hatemi-J cointegration tests. Long-run coefficients are subsequently estimated using the DOLS and FMOLS estimators, and the robustness of the empirical findings is further assessed using the ARDL approach. In addition, an interaction term is added to analyze the impact of renewable energy use on energy import-driven inflation. This interaction term is constructed to capture the extent to which renewable energy alleviates the inflationary effects of energy imports, rather than merely reflecting a generic combined effect of two variables.

The findings show that an increase in the exchange rate has an inflationary impact in Türkiye. DOLS results indicate that a one-unit increase in the exchange rate increases the inflation rate by 3.228 units. Similarly, energy imports also have a significant pressure on inflation. A one-unit increase in energy imports increases inflation by 1.642 units. Conversely, the findings indicate that renewable energy has a negative impact on inflation. DOLS results reveal that a one-unit increase in renewable energy use reduces inflation by 1.145 units. Similarly, the interaction term also has a negative impact on inflation. DOLS results indicate that a one-unit increase in the interaction term decreases inflation by 1.346 units. FMOLS results also support the DOLS results. In addition, the ARDL estimates, included as a robustness check, yield broadly consistent long-run results. In both models, renewable energy sources have a disinflationary impact. Moreover, in both models, the impact of renewable energy on mitigating inflationary pressures stemming from energy imports is stronger than the direct disinflationary impact of renewable energy. This conclusion is also reinforced by the ARDL long-run estimates, which indicate that the disinflationary role of renewable energy operates more strongly through its moderating effect on import-driven inflation.

Although this study provides important findings, it has some limitations. Future studies can address these limitations and move the study forward. First, the pressure on inflation is explained by energy imports, exchange rate and renewable energy use. In the future, factors such as global energy prices, energy supply and geopolitical risks can also be included in the model. Second, this study is limited to the case of Türkiye. In future studies, developed and developing countries can also be analyzed to compare how the results vary across different countries. Finally, the study uses data from 1980 to 2022. The consistency of the findings can be tested with longer-term data sets and different analysis techniques. These limitations may inspire future studies.

**Ethics approval and consent to participate**

Not applicable.

**Authors contribution statement**

The contribution of the 1st author to the article is 100%.

**Funding Statement**

No financial support was received from any institution for this research.

**Competing interest**

The author declares no competing interests.

## Extended Summary

## Can Renewable Energy Mitigate Inflationary Pressures from Energy Imports? Evidence from Türkiye

Renewable energy could be an important opportunity to reduce Türkiye's dependence on energy imports and alleviate inflationary pressures. The use of renewable energy sources can reduce external dependence by reducing energy imports. This in turn can alleviate pressure on the exchange rate and reduce inflationary effects. Renewable energy can also contribute to Türkiye's long-run energy strategies by improving environmental sustainability. However, more research is needed on the economic implications of this transition and its effects on inflation. Therefore, this study analyzes the potential of renewable energy to alleviate inflationary pressures from energy imports.

The study uses annual data for the period 1980-2022. This period was chosen on the basis of the availability and accessibility of the dataset for analysis. Inflation (INF) is used as the dependent variable. Energy imports (IMP), renewable energy use (REN) and exchange rate (EXC) are included in the model as independent variables. In addition, an interaction term is constructed to measure the impact of renewable energy in reducing inflationary pressures from energy imports. The interaction term is a method used in statistical modeling to examine the combined effect of two or more independent variables on the dependent variable. In this study, the interaction term measures the combined effect of these two factors in addition to the independent effects of energy imports and renewable energy use on inflation.

DOLS results show that a one unit increase in the exchange rate increases the inflation rate by 3.228 units. This result is quite significant for a foreign currency dependent economy like Türkiye. An increase in the exchange rate increases import costs. Especially in energy import-dependent countries like Türkiye, increases in exchange rates cause inflationary pressures by increasing energy costs. Since Türkiye's energy and raw material imports directly affect the general level of prices, exchange rate increases put significant pressure on inflation. Similarly, energy imports (IMP) also have an inflationary impact. According to the DOLS model, a one-unit increase in energy imports increases inflation by 1,642 units. Türkiye's energy needs are largely met by imports. These reasons increase Türkiye's sensitivity to fluctuations in global energy prices. Increases in the prices of energy resources such as oil and natural gas trigger inflation by increasing the cost of energy imports. This shows that Türkiye is vulnerable to external shocks due to its energy dependence. Renewable energy utilization (REN), on the other hand, has a downward impact on inflation. DOLS results show that a 1-unit increase in REN reduces inflation by 1.145 units. This suggests that reducing Türkiye's dependence on energy imports has a positive impact on inflation. Renewable energy sources play an important role in controlling inflation by reducing energy costs, especially through the use of domestic resources such as solar and wind energy. This result reveals the stabilizing effect of renewable energy investments on energy prices. Finally, the TERM variable refers to the interaction term. This term measures the effect of renewable energy in mitigating inflationary pressures from energy imports. According to the DOLS model, the interaction term has a negative effect. A 1-unit increase in TERM reduces the inflation rate by 1.346 units. This result suggests that renewable energy alleviates inflationary pressures stemming from energy imports. Renewable energy reduces energy costs by reducing the use of imported fossil fuels. Thus, the pressure of energy imports on prices decreases. This shows how important Türkiye's investments in renewable energy are in controlling inflation.

FMOLS results also support the DOLS results. In both models, the use of renewable energy sources has a disinflationary effect. Moreover, in both models, the effect of renewable energy in alleviating inflationary pressures stemming from energy imports is stronger than the direct disinflationary effect of renewable energy. This suggests that renewable energy investments play a critical role in containing inflation in Türkiye in the long run. Renewable energy reduces energy costs and stabilizes inflation by reducing external dependence. The robustness of these long-run findings is further confirmed by the ARDL approach, which yields broadly consistent long-run coefficients.

While this study provides important findings, it has some limitations. Future studies can address these limitations and conduct more comprehensive analyses. First, the pressure on inflation is explained by energy imports, exchange rate and renewable energy use. In the future, factors such as global energy prices, energy supply and geopolitical risks can also be included in the model. Second, this study is limited to the case of Türkiye. In future studies, developed and developing countries can also be analyzed to compare how the results vary across different countries. Finally, the study uses data from 1980 to 2022. The consistency of the findings can be tested with longer-term data sets and different analysis techniques. These limitations may give future studies the opportunity to provide more in-depth and robust results.

The results of the analysis suggest that the use of renewable energy resources can alleviate the inflationary pressures stemming from energy imports in Türkiye. In line with these findings, Türkiye's energy policy shift towards renewable energy has strategic importance for both economic performance and inflation management. Renewable energy can ease exchange rate and energy cost pressures by reducing dependence on energy imports, thereby supporting more sustainable anti-inflation strategies. Moreover, the efficient use of renewable energy sources not only enhances economic stability but also supports environmental sustainability; accordingly, a set of policy recommendations is proposed for reducing import dependence and combating inflation.

First, various supports such as tax breaks, subsidies, low-interest loans, and R&D grants are proposed to encourage renewable energy investments, alongside policies that promote domestic technology production to reduce external dependence and lower energy costs in the long run. Secondly, it is emphasized that energy efficiency projects should be prioritized and investments should be made in energy storage systems to reduce energy dependency and ensure supply security. Third, policies that encourage the use of renewable energy in energy-intensive sectors should be developed, while incentives for energy-efficient technologies can reduce industrial costs and alleviate inflationary pressures. Fourth, a mechanism that regularly monitors the relationship between energy and inflation is proposed to evaluate short- and long-term policy effects and adjust instruments when needed. Finally, it is emphasized that domestic technology production should be supported and import dependency in the energy sector should be reduced to strengthen Türkiye’s domestic energy infrastructure and mitigate inflationary pressures.

This study contributes to the literature in at least five ways: (1) This study analyzes the potential of renewable energy to reduce inflationary pressures arising from energy imports. The study aims to fill an important gap in the literature by addressing the relationship between energy imports, renewable energy and inflation. (2) Unlike other studies, this study uses the interaction term to analyze the relationship between energy imports and renewable energy use. The interaction term aims to measure the independent effects of energy imports and renewable energy use on inflation as well as the combined effect of these two factors. This allows for a more in-depth analysis. (3) Instead of traditional unit root tests, unit root tests that take structural breaks into account are also used. Specifically, the Zivot–Andrews and Lee–Strazicich tests are employed. This method allows for a better analysis of the dynamic structure of time series. It also increases the robustness of the results. (4) In addition to long-run relationships, the study also estimates long-run coefficients. These estimates contribute to a better understanding of the long-run effects of energy imports and renewable energy use. (5) DOLS and FMOLS methods are used in the analysis and the results are compared. The use of these two methods improved the accuracy of the estimation results and strengthened the reliability of the analysis. In addition, cointegration is assessed via the Johansen and Hatemi-J tests, and the robustness of the long-run evidence is further evaluated using the ARDL approach.